# White Matter Hyperintensities Segmentation Using Probabilistic TransUNet

Muhammad Noor Dwi Eldianto[1], Muhammad Febrian Rachmadi[1,2], Wisnu Jatmiko[1]

[1]*Faculty of Computer Science*, *University of Indonesia,* Depok, Indonesia
[2]Brain Image Analysis Unit, RIKEN Center for Brain Science, Wako-shi, Japan

Email: muhammad.noor02@ui.ac.id

*Abstract*—**White Matter Hyperintensities (WMH) are areas of the brain that have higher intensity than other normal brain regions on Magnetic Resonance Imaging (MRI) scans. WMH is often associated with small vessel disease in the brain, making early detection of WMH important. However, there are two common issues in the detection of WMH: high ambiguity and difficulty in detecting small WMH. In this study, we propose a method called Probabilistic TransUNet to address the precision of small object segmentation and the high ambiguity of medical images. To measure model performance, we conducted a k-fold cross validation and cross dataset robustness experiment. Based on the experiments, the addition of a probabilistic model and the use of a transformer-based approach were able to achieve better performance.**

## I. INTRODUCTION

White Matter Hyperintensities (WMH) are areas of the brain with higher intensity than other normal brain areas on Magnetic Resonance Imaging (MRI) scans [1]. WMHs are often associated with small vessel brain diseases. According to research conducted in recent decades, WMHs are a key factor in clinical outcomes in terms of cognitive and functional disorders, and are three times more likely to result in stroke and two times more likely to result in dementia [2], [3].

WMH is difficult to detect at an early stage of the disease, so many researchers have conducted research to detect WMH at an early stage of the disease. Early detection of WMH is important to provide an important opportunity for prevention and proper treatment, so that further damage to the brain does not occur. [3].

Early detection of WMH can be done by segmenting MRI images, but manually segmenting by radiologists requires a significant amount of time and cost [4]. Therefore, a system is needed that can automatically segment WMH on brain MRI images.

The advancement of technology now makes it possible to create a system that can automatically segment WMH on brain MRI images. One technology that can be used to support this is Deep Learning. A number of studies have been conducted to create a system that can automatically segment WMH using deep learning, including by [1], [4].

The study conducted by [1] added Global Spatial Information (GSI) as additional information to the input of Convolutional Neural Networks (CNN). In this study, GSI consists of the x, y, z axes and a radial filter that are encoded into the form of distance from the center of the MRI image. The MRI sequences used in this study were T1-Weighted (T1W) and Fluid-Attenuated Inversion Recovery (FLAIR). The result of adding GSI as additional information to the input of CNN is a better Dice Similarity Coefficient (DSC) value than a CNN without using GSI. This study also showed that CNN performs better than conventional machine learning methods such as Support Vector Machine (SVM) and Random Forest (RF).

In the study conducted by they attempted to investigate the robustness of Probabilistic UNet in performing WMH Segmentation. The study conducted a cross-dataset test. The datasets used in this study were MRI images from the Alzheimer's Disease Neuroimaging Initiative (ADNI) Dataset, Singapore Dataset, GE3T Dataset, and Utrecht Dataset. The cross-dataset test in this study means that a model trained using one dataset will be tested using a different dataset. This is intended to determine the robustness of the resulting model. In this study, the probabilistic model was proven to be robust and produced the highest average DSC value compared to other models such as U-Net, Attention U-Net, U-Net++, and Attention U-Net++.

In the field of semantic segmentation research, models based on Fully Convolutional Networks (FCN) are generally dominant. However, these FCN-based models have limitations in obtaining global information context. While in semantic segmentation, global information context is often important, because labeling local patches often depends on global image context [5], [6]. This has led many researchers to improve the performance of FCN-based semantic segmentation models to obtain global information context by increasing the receptive field. However, this will increase the computation weight used.

Transformer is considered a new solution to solve the problems that occur in FCN-based models. The study conducted by [5] and [6] used a Transformer-based model for segmentation. Transformer has good capabilities in obtaining global information context because Transformer is a sequence-to-sequence model that can capture contextual information.

There are various ways to apply Transformer in segmentation. The study conducted by [5] used Transformer for encoding and decoding of the original image, while [6] directly up sampled the output of Transformer. The study conducted by [7] used Transformer as an additional layer to enrich the information in the encoder. For the encoder used by [7], CNN was used to extract features from the image, obtaining low-level features to high-level features, and adding a transformer layer to enrich the information obtained.

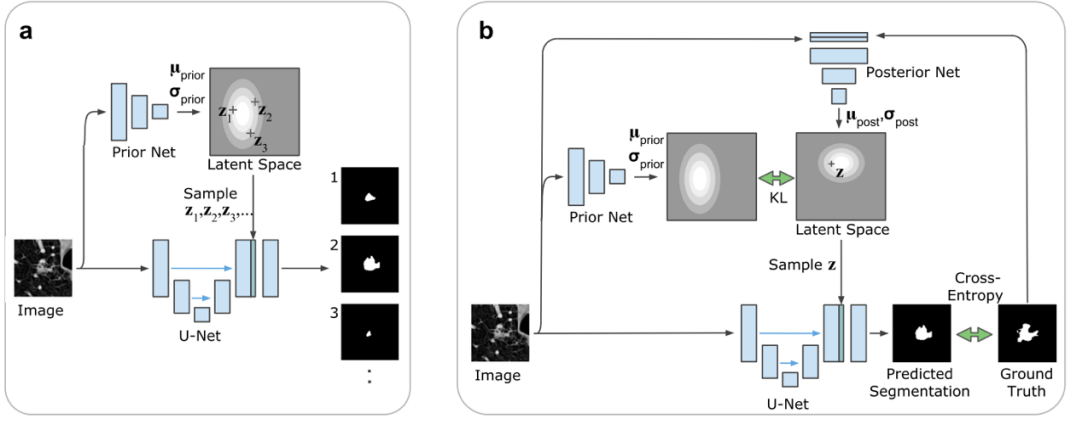

fig. 1 (a) The testing process for Probabilistic U-Net, (b) The training process for Probabilistic U-Net

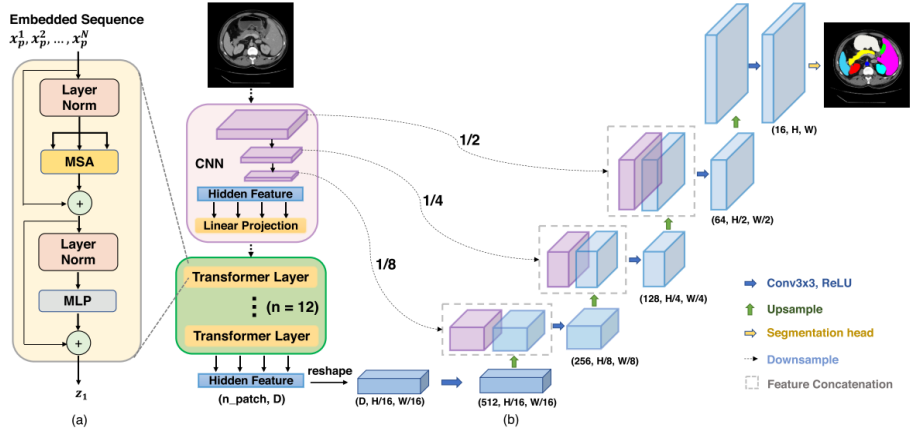

fig. 2 TransUNet architecture

In medical image segmentation research, some challenges are encountered, namely the small size of the objects to be segmented and the high ambiguity between abnormal tissue and normal objects. Therefore, to overcome this, in the study [7], the researcher used Transformer to perform precise segmentation and localization as an effort to solve the problem of segmentation on small-sized objects, while in the study conducted by [4], [8] the researcher used a probabilistic model to overcome the high ambiguity in medical images.

Based on the background above, in this research, a method will be proposed to solve both of these problems, namely Probabilistic TransUNet to overcome the precise segmentation of small-sized objects and the high ambiguity of medical images. It is expected that the resulting model will have better performance than the methods of previous research.

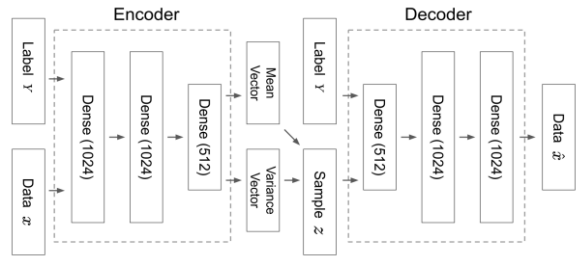

fig. 3 CVAE architecture

## II. RELATED WORK

U-Net is one of the Deep Learning architectures frequently used for Image Segmentation, especially in the medical field. U-Net consists of two important parts, the contracting path and the expansive path [9]. The contracting path in U-Net serves to obtain context, while the expansive path is useful for precise localization and image reconstruction.

Probabilistic U-Net is a development of U-Net. Probabilistic U-Net combines U-Net with Conditional Variational Auto Encoder (CVAE) [10], and through this combination, Probabilistic U-Net falls into the category of generative segmentation model. The idea of Probabilistic U-Net is to create additional features from the latent space (prior and posterior) to overcome the ambiguity of ground truth, as medical images have many possibilities in hypothesis [8].

With the addition of CVAE, the model can make hypotheses from the input image and ground truth without limitations. An illustration of the testing and training process for the probabilistic U-Net can be seen in figure 1

TABLE 1 HYPERPARAMETER CONFIGURATION

| Model | Hyperparameter |
|---|---|
| UNet | Filter = 64, 128, 256, 512, 1024 |
| Probabilistic UNet | Filter = 32, 64, 128, 256, 512 |
|  | Latent dim = 6 |
| TransUNet | Backbone = ResNet50V2 |
|  | Filter = 16, 32, 64, 128 |
|  | Transformer layer = 12 |
| Probabilistic TransUNet | Backbone = ResNet50V2 |
|  | Filter = 16, 32, 64, 128 |
|  | Transformer layer = 12 |
|  | Latent dim = 6 |

The CVAE (Conditional Variational Autoencoder) is utilized as an architecture that can provide additional information in the final process of the U-Net architecture, as demonstrated in the study conducted by [8] and based on Figure 1b. The input image is inputted into the CVAE, which is comprised of the Prior Net and Posterior Net. An illustration of the CVAE can be seen in figure 3.

TransUNet is an advanced development of the widely used U-Net architecture for medical image segmentation. TransUNet aims to provide more information from the global context generated by the CNN to achieve more precise localization, by tokenizing image patches from the feature map generated by the CNN [7].

This more precise localization can improve the performance of U-Net in medical image segmentation, as the objects being segmented are typically very small in size. TransUNet architecture can be seen in figure 2.

TABLE 2 *DATASET*

| Dataset | Amount | Usability |
|---|---|---|
| Dataset ADNI | 840 | Training |
| Dataset Singapore | 460 | *cross dataset robustness* |
| Dataset GE3T | 475 | *cross dataset robustness* |
| Dataset Utrecht | 439 | *cross dataset robustness* |

A study conducted by [7] also used a model based on the Transformer for segmentation. [7] used the Transformer as an encoder for abdominal CT scan image segmentation. Unlike [5] which directly used the Transformer for encoding and decoding the original image and which directly upsampled the output of the Transformer, [7] used a CNN in the encoder to extract features from the image, resulting in low level to high level features, and added a transformer layer to obtain global contextual information.

assumes that global context alone is not sufficient because good upsampling requires low level to high level features. Therefore, to not lose this information, [7] uses a CNN for feature extraction. The CNN is also used to reduce the size of the original image to reduce computational burden. After performing feature extraction with the CNN, the output from the CNN is subjected to linear projection and patch embedding. The output from the Transformer layer, which carries global contextual information, is added as additional information to the output from the CNN and upsampled to become the segmented image.

Based on previous literature studies, there has been no research using transformer-based architectures for segmenting WMH images and no research combining probabilistic models with TransUNet.

Therefore, in this research, a method will be proposed to address these two problems, namely Probabilistic TransUNet for accurately segmenting small objects and addressing the high medical image ambiguity. It is expected that the resulting model will have better performance than previous methods based on the DSC values obtained from cross-dataset robustness experiments and K-fold cross validation as conducted by [4], [11], [12] .

## III. METHOD

The research process begins by formulating the research problem based on the research background. Then, literature study is conducted to solve the problems identified in the research problem. The research method is then designed based on the conclusions obtained from the literature study. Next, datasets are collected for WMH segmentation. There are two datasets used in this research, which will subsequently be used for training and evaluating cross-dataset robustness. Subsequently, preliminary research is conducted to determine whether the TransUNet model can be used for WMH segmentation and to see the initial performance of the model, as well as trying to combine the Probabilistic model and TransUNet to identify the challenges that arise from the combination of the two models.

The combination of the two models, the Probabilistic and the TransUNet, will subsequently be referred to as the Probabilistic TransUNet, which is a new proposed model. Next, the performance of the proposed model will be tested. Performance testing is based on the DSC score obtained from cross-dataset robustness, which will be compared to previous methods. The results of the performance testing will be analyzed, and conclusions will be drawn.

### A. Dataset

The dataset used in this research is brain MRI images from the Alzheimer's Disease Neuroimaging Initiative (ADNI) [13] as the training data and the White Matter Hyperintensities Challenge [14] as the testing data. The ADNI dataset is chosen because it has been used in research, such as that conducted by [1] and [4] while the White Matter Hyperintensities Challenge dataset has been used as testing data by [4] to assess the robustness of the built model.

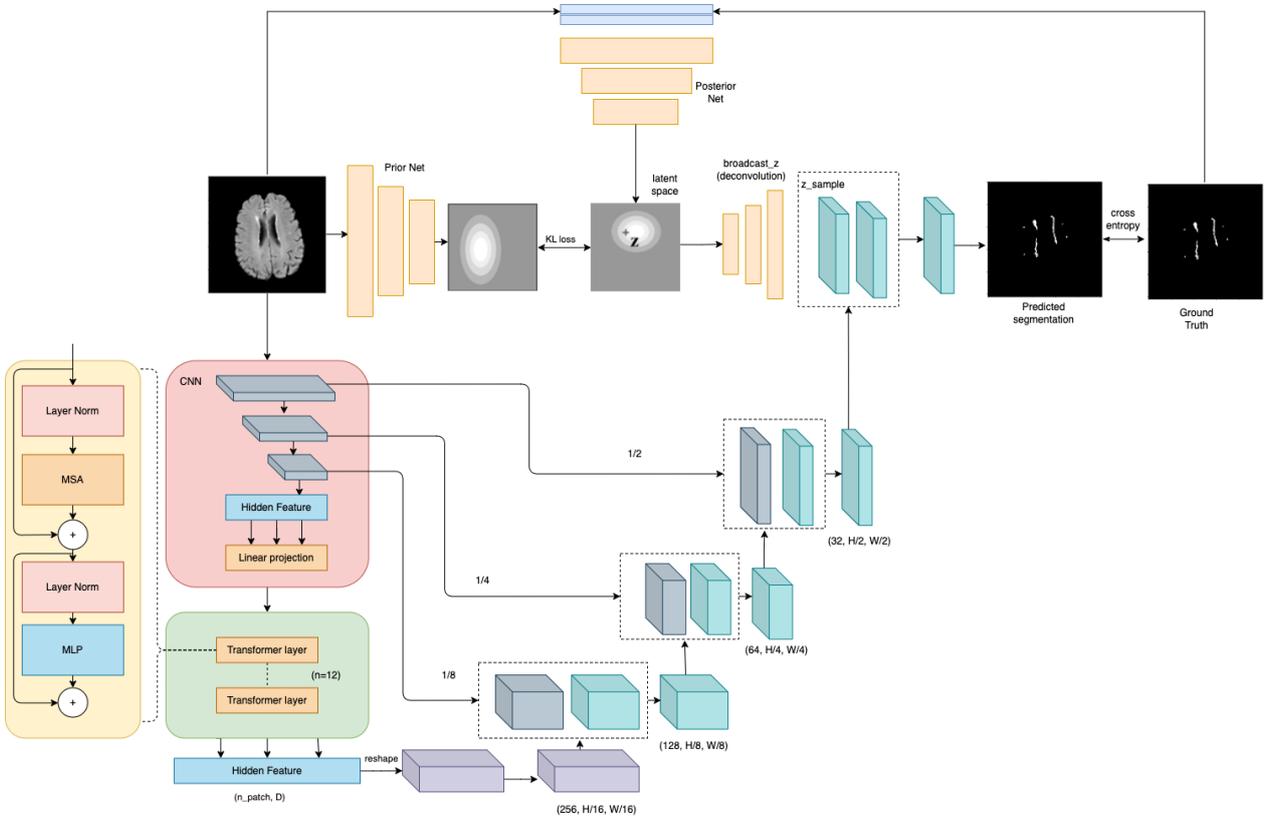

fig. 4 Probabilistic TransUNet (proposed) architecture

The two datasets have different characteristics. The ADNI dataset consists of 20 patients, and each patient has 3 MRI images taken at different times, resulting in a total of 60 MRI images in the ADNI dataset, with each image having dimensions of 256 x 256 x 35 pixels. On the other hand, the White Matter Hyperintensities Challenge dataset consists of 3 different institutions (Singapore, GE3T, and Utrecht), each of which has 20 MRI images from 20 patients, resulting in a total of 60 MRI images in the White Matter Hyperintensities Challenge dataset, with the dimensions of each image from each institution being 232 x 256 x 48, 132 x 256 x 83, and 132 x 256 x 48, respectively.

In all datasets, data augmentation will be performed in the form of horizontal flip, rotate, Z-Score Normalization, and rescale and zero padding to make them all have the same size of 128 x 128 pixels. The rescale is done to reduce computational load

After performing data augmentation, the following amounts were obtained for each dataset, which will subsequently be used for training and evaluating cross-dataset robustness. Datasets composition can be seen in table 2:

*B. Proposed Archietcture*

In this research, the researcher proposes a new model called the Probabilistic TransUNet. This model is a modification of the Probabilistic UNet in the research conducted by [8]. The modification made is by replacing the UNet with the TransUNet architecture from [7].

The use of the probabilistic model aims to address the ambiguity problem that occurs in medical image segmentation research, while replacing the UNet with the TransUNet aims to achieve more precise localization.

TABLE 3 HYPERPARAMETER GENERAL CONFIGIURATION

| Hyperparameter | Configuration |
|---|---|
| Epoch | 500 |
| Learning rate | 0.001 |
| Optimizer | Adam |
| Batch size | 8 |

In the preliminary study, the researcher encountered difficulties in combining the probabilistic model with the TransUNet. The results obtained from the Probabilistic TransUNet, such as producing repeated pattern noise. After further investigation of the code created, the process of combining the output of the probabilistic model with the TransUNet, if referring to the research [8] uses the tf.tile() function. The tf.tile() function repeats the input to achieve a certain size, with the aim of equalizing the dimensions between the output of the probabilistic model and the TransUNet. The merging process requires the same dimensions between the output of the probabilistic model and the TransUNet.

To solve this problem, the researcher made changes to the merging function by performing deconvolution with a stride size of 2 so that the output of the probabilistic model has the same size as the output of the TransUNet. This can solve the problems the researcher encountered in the experiment. A description of the Probabilistic TransUNet architecture proposed by the researcher can be seen in the figure 4.

*C. Experiment Scenario*

Broadly speaking, the experimental scenario carried out in this research will follow the experimental scenario carried out by [4] which is using K-fold cross validation and evaluating cross-dataset robustness, but with a slight adjustment to the configuration of the hyperparameters used.

TABLE 4 TRAINING TIME PER EPOCH (SECONDS)

| Model | Training time per epoch (seconds) |
|---|---|
| UNet [9] | 4 |
| Probabilistic UNet [8] | 3 |
| TransUNet [7] | 42 |
| Probabilistic TransUNet (proposed) | 42 |

TABLE 5 K-FOLD CROSS VALIDATION RESULT

| Model | DSC score (std) |
|---|---|
| UNet [9] | 0.612 (0.062) |
| Probabilistic UNet [8] | 0.526 (0.113) |
| TransUNet [7] | 0.684 (0.063) |
| Probabilistic TransUNet (proposed) | **0.742 (0.024)** |

In the K-fold cross validation scenario, the number of k used is 5, so the number of ADNI datasets used for the training process is 672 images and 168 images for the testing process in each fold run. As for the changes in the configuration of the hyperparameters used, they will be further explained in the experimental setup section. In the training process, the saved weights are the weights that achieve the highest DSC validation score.

The experiment in this research used the Python programming language with the help of the Tensorflow 2.5.0 library, which was run on an Nvidia DGX-1 computer, an Nvidia Tesla V-100 GPU with 32 GB of memory.

*D. Experimental Setup*

The configuration of the model used in the training process for all architectures uses the same epoch, learning rate, optimizer, and batch size. The hyperparameter configuration for each model can be seen in the table 1 and table 3.

*E. Evaluation Method*

The evaluation to be performed on the built model will use the Dice Similarity Coefficient (DSC). DSC is chosen because it is more representative in evaluating Image Segmentation. If Accuracy is used, the results will not be representative due to the large imbalance of classes in Medical Image Segmentation. The equation for DSC is as follows:

$$DSC = \frac{2 \cdot TP}{2 \cdot TP + FP + FN}$$

Where:
- TP = True Positive
- FP = False Positive
- TN = True Negative
- FN = False Negative

## IV. RESULTS AND DISCUSSION

In this section, the results of the experiments carried out will be elaborated upon and the proposed model will be compared with the comparative model. The analysis will be conducted to determine the strengths and weaknesses of each model and to find out which model is better.

TABLE 6 CROSS DATASET ROBUSTNESS RESULT

| Model | DSC (std) | | | Average DSC (std) |
|---|---|---|---|---|
| | **Singapore** | **GE3T** | **Utrecht** | |
| UNet [9] | 0.552 (0.340) | 0.626 (0.316) | 0.578 (0.312) | 0.585 (0.322) |
| Probabilistic UNet [8] | 0.553 (0.357) | 0.670 (0.330) | 0.581 (0.330) | 0.601 (0.339) |
| TransUNet [7] | **0.569 (0.357)** | 0.663 (0.311) | **0.599 (0.322)** | **0.610 (0.330)** |
| Probabilistic TransUNet (proposed) | 0.562 (0.357) | **0.680 (0.311)** | 0.574 (0.331) | 0.605 (0.333) |

### A. Comparison based on Training Time

In this section, the researcher compares the training time of each model run using an Nvidia DGX-1 computer. The training time of each model can be seen in table 4:

As can be seen in table 4, the difference in training time required for 1 epoch on the TransUNet takes about 10 times longer than the UNet, but when compared to the Probabilistic TransUNet, the time required is the same, which is 42 seconds in 1 epoch. Therefore, it can be said that the addition of the probabilistic model to the TransUNet does not cause an increase in training time compared to the TransUNet.

If we compare the UNet with the Probabilistic UNet, the Probabilistic UNet requires a shorter time, namely 3 seconds, due to differences in the configuration of the number of filters. The number of filters used in the Probabilistic UNet [8] at each level in sequence is (32, 64, 128, 256, 512), while the number of filters used by the UNet [9] is (64, 128, 256, 512, 1024), which is what causes the time required by the Probabilistic UNet in 1 epoch to be less than the UNet. The researcher has also tried to modify the number of filters used by the Probabilistic UNet to have the same number of filters as the UNet, which made the obtained DSC score tend to decrease and the training time in one epoch became three times longer, that is, 9 seconds.

### B. Comparison based on K-Fold Cross Validation

In this section, the researcher compares the results obtained by each model. The model with the highest DSC score is the model with the best performance in the K-fold cross validation experiment and will be marked with bold print. The results obtained in the K-fold cross validation experiment can be seen in the table 5.

Based on table 5, the Probabilistic TransUNet has the highest DSC score. The Probabilistic TransUNet has a DSC score of 0.742 with a standard deviation of 0.024.

### C. Comparison based on Cross Dataset

In this section, the researcher compares the results obtained by each model when evaluated in the cross-dataset. The model that has been trained with the ADNI data will be tested using 3 different datasets: Singapore, GE3T, and Utrecht datasets. The model with the highest DSC score is the model with the best performance in the cross-dataset experiment and will be marked with bold print. The results obtained in the cross-dataset experiment can be seen in table 6

### D. Discussion

In this section, the researcher will analyze the results obtained. The analysis of the results will be divided into two parts: based on the DSC scores obtained by each model and based on the visualization of the model's ability to segment small WMH objects.

*1) Based on DSC Score*

Based on the results of the K-fold cross validation experiment shown in table 5, the Probabilistic TransUNet obtained the highest DSC score with a score of 0.742 (0.024), an increase of about 0.058 compared to the TransUNet.

Based on the results of the cross-dataset evaluation experiment, referring to table 6, TransUNet is generally the best model with an average DSC value of 0.610. When looking at the DSC scores obtained for each dataset, TransUNet obtained the highest DSC score in almost every dataset. TransUNet is only outperformed by the Probabilistic TransUNet in the GE3T dataset. In general, the use of transformer-based models (TransUNet and Probabilistic TransUNet) in the cross-dataset evaluation experiment outperforms CNN-based models (UNet and Probabilistic UNet).

The addition of a probabilistic model to TransUNet does not seem to significantly increase the DSC score, with only a 0.017 increase in the DSC score in the GE3T dataset. On the other hand, the addition of a probabilistic model to UNet resulted in an increase in the DSC score in all datasets. The increase in the DSC score from the addition of a probabilistic model can be seen in table 7:

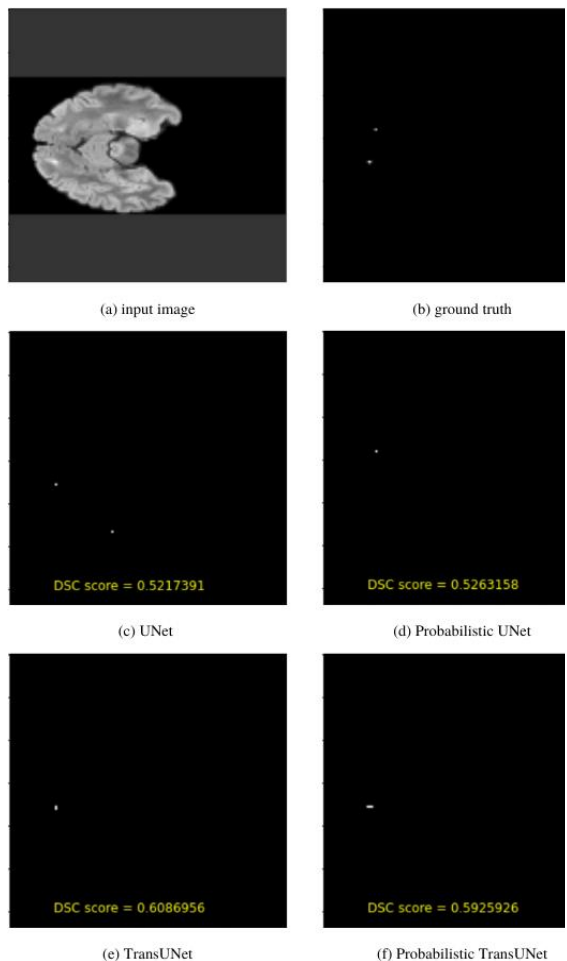

fig. 5 The segmentation results from all the models

Based on table 7, the addition of a probabilistic model to both CNN-based (UNet) and transformer-based (TransUNet) models results in an average increase in the DSC score.

*2) Based on the Ability to Perform Small Object Segmentation*

In this section, the researcher will provide an analysis of the experimental results based on the ability of the models to perform small object segmentation. The results of the segmentation of each model for small-sized objects can be seen in Figure 5.

Based on Figure 5, transformer-based models (TransUNet and Probabilistic TransUNet) have a higher DSC score than CNN-based models (UNet and Probabilistic UNet). Although the segmentation results of all models are still undersegmented and oversegmented, it can be said that transformer-based models could perform more precise segmentation than CNN-based models based on these results.

V. CONCLUSION AND FUTURE WORK

In this study, the use of Transformer-based models (TransUNet and Probabilistic TransUNet) was found to provide more precise and better segmentation results, as demonstrated by the higher DSC scores obtained compared to CNN-based models (UNet and Probabilistic UNet) and their ability to segment small WMH objects.

The addition of a Probabilistic model was able to improve the performance of both CNN-based and Transformer-based models, although the increase in Transformer-based models was relatively smaller compared to the addition of a Probabilistic model in CNN-based models.

For future research, it is suggested to conduct experiments with a variety of different filter sizes in both CNN-based and Transformer-based models, with the hope of producing more varied features and improving performance. Additionally, separating the dataset based on its characteristics can allow for more detailed experiments on the models, enabling the identification of which model is more suitable for segmentation on datasets with specific characteristics.

TABLE 7 INCREASE IN DSC SCORE

| Model | Dataset | | | Average |
|---|---|---|---|---|
| | Singapore | GE3T | Utrecht | |
| UNet [9] | +0.001 | +0.044 | +0.003 | +0.016 |
| TransUNet [7] | -0.007 | +0.017 | -0.005 | +0.001 |

ACKNOWLEDGMENT

We would like to express our gratitude to the Tokopedia- UI AI Center and the Faculty of Computer Science at the University of Indonesia for providing us with the NVIDIA DGX-1, which we used to conduct our experiments. We would also like to express our gratitude to Yingkai Sha for creating the keras-unet-collection library (https://github.com/yingkaisha/keras-unet-collection), which we used in this research.

REFERENCES

[1] M. F. Rachmadi, M. del C. Valdés-Hernández, M. L. F. Agan, C. di Perri, and T. Komura, "Segmentation of white matter hyperintensities using convolutional neural networks with global spatial information in routine clinical brain MRI with none or mild vascular pathology," *Computerized Medical Imaging and Graphics*, vol. 66, pp. 28–43, 2018, doi: 10.1016/j.compmedimag.2018.02.002.

[2] N. D. Prins and P. Scheltens, "White matter hyperintensities, cognitive impairment and dementia: An update," *Nat Rev Neurol*, vol. 11, no. 3, pp. 157–165, 2015, doi: 10.1038/nrneurol.2015.10.

[3] J. M. Wardlaw, M. C. Valdés Hernández, and S. Muñoz-Maniega, "What are white matter hyperintensities made of? Relevance to vascular cognitive impairment," *J Am Heart Assoc*, vol. 4, no. 6, p. 001140, 2015, doi: 10.1161/JAHA.114.001140.

[4] R. Maulana, "Robustness of Probabilistic U-Net for Automated Segmentation of White Matter Hyperintensities in Different Datasets of Brain MRI," pp. 1–7.

[5] R. Strudel, R. Garcia, I. Laptev Inria, and C. Schmid Inria, "Segmenter: Transformer for Semantic Segmentation," *arXiv preprint*, pp. 7262–7272, 2021, [Online]. Available: https://github.com/rstrudel/segmenter

[6] S. Zheng *et al.*, "Rethinking Semantic Segmentation from a Sequence-to-Sequence Perspective with Transformers," pp. 6877–6886, 2021, doi: 10.1109/cvpr46437.2021.00681.

[7] J. Chen *et al.*, "TransUNet: Transformers Make Strong Encoders for Medical Image Segmentation," pp. 1–13, 2021, [Online]. Available: http://arxiv.org/abs/2102.04306

[8] S. A. A. Kohl *et al.*, "A probabilistic U-net for segmentation of ambiguous images," *Adv Neural Inf Process Syst*, vol. 2018-Decem, no. NeurIPS, pp. 6965–6975, 2018.

[9] O. Ronneberger, P. Fischer, and T. Brox, "U-Net: Convolutional Networks for Biomedical Image Segmentation," *ArXiv*, pp. 1–8, 2015, [Online]. Available: http://lmb.informatik.uni-freiburg.de/%0Aarxiv:1505.04597v1

[10] C.-Y. Tsao, J.-H. Chen, S. Y.-C. Chen, and Y.-C. Tsai, "Data Augmentation for Deep Candlestick Learner," no. May, 2020, [Online]. Available: http://arxiv.org/abs/2005.06731

[11] S. Yadav and S. Shukla, "Analysis of k-Fold Cross-Validation over Hold-Out Validation on Colossal Datasets for Quality Classification," *Proceedings - 6th International Advanced Computing Conference, IACC 2016*, no. Cv, pp. 78–83, 2016, doi: 10.1109/IACC.2016.25.

[12] N. Furuhashi, S. Okuhata, and T. Kobayashi, "A robust and accurate deep-learning-based method for the segmentation of subcortical brain: Cross-dataset evaluation of generalization performance," *Magnetic Resonance in Medical Sciences*, vol. 20, no. 2, pp. 166–174, 2021, doi: 10.2463/mrms.mp.2019-0199.

[13] S. G. Mueller *et al.*, "The Alzheimer's disease neuroimaging initiative," *Neuroimaging Clin N Am*, vol. 15, no. 4, pp. 869–877, 2005, doi: 10.1016/j.nic.2005.09.008.

[14] H. J. Kuijf *et al.*, "Standardized Assessment of Automatic Segmentation of White Matter Hyperintensities and Results of the WMH Segmentation Challenge," *IEEE Trans Med Imaging*, vol. 38, no. 11, pp. 2556–2568, 2019, doi: 10.1109/TMI.2019.2905770.